\documentstyle[aps,amsmath,amssymb,preprint]{revtex}
\title{Vacuum Polarization of Massive Scalar Fields \\
on the  Black  Hole Horizon}
\author{Hiroko Koyama, Yasusada Nambu and Akira  Tomimatsu}
\date{\today}
\address{Department of Physics, Graduate
  School of Science, Nagoya University, Nagoya 464-8602, Japan}

\def\ep{\epsilon}
\def\al{\alpha}
\tighten
\begin{document}
\preprint{DPNU 00-11}
\draft
\maketitle
\begin{abstract}

  Vacuum polarization of massive scalar fields in a thermal state at
  arbitrary temperature is studied near the horizon of a
  Reissner-Nordstr\"{o}m black hole.  We derived an analytic form of
  $\langle\phi^2\rangle$ approximately in the large mass limit near the
black hole horizon. We uses the zeroth order WKB approximation and power
series expansion near the horizon for the Euclideanized mode function. Our
  formula for the vacuum polarization shows regular behavior on the horizon
if the temperature of
  the scalar field is equal to the Hawking temperature of the black
  hole.  The finite part of the vacuum polarization agrees with the
  result of the DeWitt-Schwinger approximation up to $O(m^{-4})$ which
  is the next leading order of the expansion.

\end{abstract}

\section{Introduction}    
Since Hawking's discovery of black holes radiation \cite{Hawking2}, it
has been an important subject to investigate quantum effects in black
hole spacetimes and a great deal of effort has been done. The
expectation values of the regularized vacuum polarization $ \langle
\phi ^2 \rangle $ and the stress energy tensor $ \langle T_{\mu \nu}
\rangle $ are important quantities to understand quantum effects in
black hole spacetimes.  They represent the effect of quantum
fluctuations and back reaction on black hole spacetimes and play an
important
role in the context of semi-classical theory of quantum gravity.

For a scalar field in the large mass limit, of which Compton length is
much smaller than the characteristic curvature scale of background
geometry, we can evaluate
regularized $ \langle \phi ^2 \rangle $ and $ \langle T_{\mu \nu}
\rangle $ by using the DeWitt-Schwinger
approximation\cite{Christensen,DeWitt,Frolov,Frolov2,Parker}.  This is an
adiabatic expansion in terms of inverse power series of the mass of scalar
field.  One can evaluate  divergent and finite parts of $ \langle
\phi ^2 \rangle $ and $ \langle T_{\mu \nu} \rangle $ up to arbitrary
order.  Frolov and Zelnikov \cite{Frolov,Frolov2} suggest that the
finite parts of $ \langle \phi ^2 \rangle $ and $ \langle T_{\mu \nu}
\rangle $ calculated by the DeWitt-Schwinger expansion coincides with
the value calculated by assuming the Hartle-Hawking state\cite{Hartle} for
the scalar field, although the DeWitt-Schwinger expansion does not impose
any condition on the state of the scalar field.
The justification is confirmed numerically near the
horizon\cite{Anderson1,Anderson3}.

The Hartle-Hawking state is defined by requiring that the Euclidean
Green's function is regular on the horizon\cite{Hartle}. Its physical
meaning is
that a black hole is placed in a cavity and is equilibrium with
thermal black-body radiation.  The thermal state is defined by the
Euclidean Green's function by imposing  periodicity in the imaginary
time coordinate with a period of the inverse of the temperature.  On the
order hand,   in the DeWitt-Schwinger expansion, the condition which
characterizes the
thermal state is not imposed.  So it is not obvious whether the
DeWitt-Schwinger expansion yields the same finite part of the
Hartle-Hawking state near the horizon.

In this paper, we consider scalar fields in  static spherically
symmetric background spacetimes. By assuming that the mass of scalar
fields is large, we calculate $ \langle \phi ^2 \rangle $ in the
thermal state with arbitrary temperature using point splitting method
near the horizon. A similar calculation is already done by Anderson {\it et
al.}\cite{Anderson1,Anderson3}, but their expression of the vacuum
polarization  diverges on the horizon which contradicts with their numerical
calculation.  Our purpose is to derive the vacuum polarization which is
regular on the horizon, and resolve why the finite part of $
\langle \phi ^2 \rangle $ in the Hartle-Hawking state near the horizon
can be approximated by the DeWitt-Schwinger expansion which does not
assume any condition on the thermal state.

The plan of this paper is as follows. In Sec.II, we review the method to
calculate $\langle \phi ^2
\rangle $ in a thermal state using point splitting regularization.  In
Sec.III, we present our method to calculate $\langle \phi ^2 \rangle $
in the large mass limit near the black hole horizon.  Sec.IV is devoted to
conclusion and discussion.  Throughout the paper, we use units such
that $\hbar =c= G=k_B=1$.  Our sign conventions are those of Misner,
Thorne and Wheeler \cite{MTW}.

\section{derivation of vacuum polarization $\langle\phi^2\rangle$ in a
thermal state}

We consider a minimally coupled scalar field with mass $m$ in a
thermal state at arbitrary temperature $T$ in a
Reissner-Nordstr\"{o}m spacetime. The metric in Euclidean section
$\tau =-it$ is given by
\begin{equation}
  \label{eq:metric}
  ds^2 =fd\tau ^2 +\frac{1}{f}dr^2+r^2d\Omega ^2,
\end{equation}
where $f=(r-r_+)(r-r_-)/r^2$, $r_+$ and $r_-$ are the location of
inner and outer horizon, respectively. The surface gravity of a black
hole is given by $\kappa = (r_+-r_-)/(2r_+^2)$. The vacuum polarization
$\langle \phi ^2\rangle $ can be computed from the Euclidean Green's
function by noting the identity
\begin{equation}
  \label{eq:unregphi2}
  \langle \phi ^2 \rangle _{\text{unreg}} =
  \frac{1}{2}G^{(1)}(x,x)=iG_{F}(x,x)=G_{E}(x,x),
\end{equation}
where $G^{(1)}$ is the Hadamard Green's function, $G_{F}$ is the
Feynman Green's function, and $G_{E}$ is the Euclidean Green's
function, respectively.  $G_{E}$ is divergent at coincident limit and
must be regularized.  In this paper, the covariant point-splitting
method is employed for the regularization of ultraviolet divergences.
We start from an expression for $G_{E}(x,x')$ when the points $x$ and
$x'$ are separated. Next we prepare an appropriate counterterm to
subtract the divergence, and then take coincident limit. This procedure
is shown schematically as
\begin{equation}
  \label{eq:}
  \langle \phi ^2\rangle _{\text{reg}}= \lim _{x' \to x} \left[
    G_{E}(x,x')-G_{DS}(x,x')\right],
\end{equation}
where $G_{DS}(x,x')$ is a point splitting counterterm needed to
regularize $G_{E}(x,x')$. The Euclidean Green's function obeys the
equation
\begin{equation}
  \label{eq:2.4}
  \left[\square -m^2 \right]G_{E}(x,x')= -\frac{\delta
    ^4(x,x')}{g^{1/2}(x)},
\end{equation}
where $m$ is the mass of the scalar field and d'Alembertian $\Box $
evaluated using the Euclidean metric (\ref{eq:metric}).  The point splitting
counter term needed to renormalize $G_E(x,x')$ in arbitrary spacetime
is\cite{Christensen}
\begin{equation}
\begin{split}
  \label{eq:DeWitt-Schwinger}
  G_{DS}(x,x') &=\frac{1}{8\pi ^2 \sigma}+\frac{m^2+(\xi -
    \frac{1}{6})R}{8\pi ^2 } \left[\gamma +\frac{1}{2}\ln
    \left[\frac{m^2 \sigma}{2}\right]\right]  \\
  &\qquad -\frac{m^2}{16\pi ^2}+\frac{1}{96 \pi ^2}R_{\alpha \beta}
  \frac{\sigma ^{,\alpha}\sigma ^{,\beta}}{\sigma},
\end{split}
\end{equation}
where $\xi$ is coupling constant to the scalar curvature $R$, $\sigma$
is the one half the square of geodesic distance between the points $x$
and $x'$ and $\gamma$ is  Euler's constant.

For simplicity of calculation, the quantity $\langle \phi ^2 \rangle$
is evaluated by timelike point separation $\tau ' \neq \tau $.  By assuming
that the  Green's function is periodic in $\ep=\tau-\tau'$ with period
$1/T$, the
expression for the vacuum polarization in the thermal state at
temperature $T$ is given by
\begin{align}
  \label{eq:phi2}
  \langle \phi ^2 \rangle _{T} &= \lim _{\tau' \to \tau}\left[
    G_{E}(x,\tau;x,\tau ')- G_{DS}(x,\tau;x,\tau ')\right]\notag\\
  &= \lim _{\tau' \to \tau}\Bigg[ \frac{T}{4\pi}\sum _{l=0}^{\infty}
  \left[(2l+1)p_{0l}q_{0l}-\frac{1}{r\sqrt{f}}\right] \notag\\
  &\qquad
  +\frac{T}{2\pi}\sum _{n=1}^{\infty} \cos [n 2\pi T(\tau- \tau')]
  \sum _{l=0}^{\infty} \left[(2l+1)p_{nl}q_{nl}- \frac{1}{r\sqrt{f}}
  \right]- G_{DS}(x,\tau;x,\tau ')\Bigg] \notag \\
 &\equiv \langle\phi^2\rangle_{n=0}+\langle\phi^2\rangle_{n\ge 1}-
   \langle\phi^2\rangle_{DS},
\end{align}
where
\begin{equation}
  \label{eq:DS}
\langle\phi^2\rangle_{DS}=\frac{m^2}{8\pi^2}\left(\ln\frac{m\sqrt{f}\ep}{2}+
\gamma -\frac{1}{2}\right)-\frac{1}{4\pi^2f\ep^2}+
 \left(\frac{f'{}^2}{192\pi^2
f}-\frac{f''}{96\pi^2}-\frac{f'}{48\pi^2r}\right).
\end{equation}
The term $ 1/(r\sqrt{f})$ is subtracted to cancel the superficial
divergence which comes from the choice of timelike point-splitting.
$p_{nl}$ and $q_{nl}$ are independent solutions of the mode equation
\begin{eqnarray}
  \label{eq:modeeq}
  \frac{d^2\chi _{nl}}{dr^2}
  +\left[\frac{2}{r}+\frac{1}{f}\frac{df}{dr}\right]\frac{d\chi _
    {nl}}{dr} -\left[\frac{(2\pi n
      T)^2}{f^2}+\frac{l(l+1)}{r^2f}+\frac{m^2}{f} \right]\chi _{nl}
  =0
\end{eqnarray}
and satisfy the Wronskian condition
\begin{equation}
  \label{eq:Wronskian}
  p_{nl}\frac{dq_{nl}}{dr}- q_{nl}\frac{dp_{nl}}{dr} =-\frac{1}{r^2f}.
\end{equation}

Anderson {\it et al.}\cite{Anderson1,Anderson3} calculated the second
and the third term of (\ref{eq:phi2}) analytically using the second order
WKB
approximation . Their result is
\begin{equation}
\begin{split}
  \label{eq:nge 1mode}
  \langle \phi ^2 \rangle_{n\ge 1}-\langle\phi^2\rangle_{DS} &= -
  \frac{m^2}{16\pi ^2}\ln \left(\frac{m^2 f}{16 \pi ^2 T^2} \right)
  +\frac{m^2}{16\pi ^2} \\ &\qquad+\frac{T^2}{12f} -
  \frac{{f'}^2}{192\pi^2 f}-\frac{f''}{96\pi ^2}-\frac{f'}{48\pi ^2r}.
\end{split}
\end{equation}
The expression (\ref{eq:nge 1mode}) was obtained by taking into account
only the contribution from $n\ge 1$ mode.  In the case of massless
scalar fields, excluding the lowest frequency mode $n=0$ corresponds
to imposing an infrared cutoff.  The expression (\ref{eq:nge 1mode})
gives good approximation for massless fields both near and far from
the horizon in Schwarzschild and Reissner-Nordstr\"{o}m
spacetimes. This was confirmed by their numerical
calculation\cite{Anderson3}.

When the result (\ref{eq:nge 1mode}) is applied to the massive scalar
field, however, two unpleasant features appear.  Firstly, the
regularized expression always has logarithmic divergence on the
horizon $f=0$ even if the scalar field is in the Hartle-Hawking state.
It is not consistent with their numerical work which shows regular
behavior of $\langle\phi^2\rangle_T$ on the horizon.  Secondly, the
finite terms of $O(m^2)$ and $O(m^0)$ are left in the regularized
expression.  This disagrees with the result of the DeWitt-Schwinger
approximation by Frolov\cite{Frolov} which does not contain such
finite terms.  Therefore the expression (\ref{eq:nge 1mode}) cannot yield
good approximation for the massive scalar field. In the next section, we
present the method to improve their approximations and resolve these
problems.

\section{vacuum polarization near the horizon in the large  mass limit }
In this section, we present the method to resolve the unwanted
behavior arose in the expression of $\langle\phi^2\rangle_{T}$ near
the horizon $f\sim 0$. We assume that the mass $m$ of the scalar field
is large:
\begin{equation}
  m \gg \frac{1}{r_{+}}.
\end{equation}
We evaluate $n\ge 1$ and $n=0$ contribution to $\langle\phi^2\rangle_T$
separately.

\subsection{$n\ge1$ contribution}

We can use the WKB approximation for $n\ge 1$ mode. The zeroth order
WKB solution is given by
\begin{align}
  &p_n=\frac{1}{\sqrt{2r^2W}}\exp\left[\int dr\frac{W}{f}\right],\quad
   q_n=\frac{1}{\sqrt{2r^2W}}\exp\left[-\int dr\frac{W}{f}\right],
   \notag \\
   &;\qquad
   W^2=\kappa^2n^2+m^2f+\frac{\left(l+\frac{1}{2}\right)^2}{r^2}f
   =\left(\alpha_n^2+\frac{\left(l+\frac{1}{2}\right)^2}{r^2}\right)f
     ,\\
   &\qquad\alpha_n^2=m^2+\frac{\kappa^2n^2}{f},\quad \kappa=2\pi T.\notag
\end{align}
The WKB approximation is correct near the horizon. Using this solution, we
have
\begin{equation}
  \langle\phi^2\rangle_{n\ge
    1}=\frac{T}{2\pi}\sum_{n=1}^{\infty}\cos(n\kappa\ep) I_n,
\end{equation}
where
\begin{align}
  I_n&\equiv
  \sum_{l=0}^{\infty}\left[(2l+1)p_nq_n-\frac{1}{r\sqrt{f}}\right]
  \notag \\
    &=\frac{1}{r\sqrt{f}}\sum_{l=0}^{\infty}\left\{
\frac{l+\frac{1}{2}}{\left[r^2\alpha_n^2+\left(l+\frac{1}{2}\right)^2\right]
^{1/2}}-1\right\}.
\end{align}
$I_n$ can be evaluated using the Plana sum formula\cite{Anderson1,Anderson3}
under the
assumption $f\sim0, \al_n\sim\kappa n/f$:
\begin{align}
\sum_{l=0}^{\infty}\frac{l+\frac{1}{2}}{\left[r^2\alpha^2_n+\left(l+\frac{1}
{2}\right)\right]^{1/2}}&=
\frac{1}{2}\frac{\frac{1}{2}}{\left[r^2\alpha_n^2+\frac{1}{4}\right]^{1/2}}+
    \int_0^{\infty}dl\frac{l+\frac{1}{2}}{\left[r^2\alpha_n^2
     +\left(l+\frac{1}{2}\right)^2\right]^{1/2}}
    \notag \\
 & \qquad +i\int_0^\infty\frac{dl}{e^{2\pi l}-1}
\left\{\frac{il+\frac{1}{2}}{\left[r^2\alpha_n^2+\left(il+\frac{1}{2}\right)
^2\right]^{1/2}}-(l\rightarrow -l)\right\} \notag \\
 &=\left.\left[\left(l+\frac{1}{2}\right)^2+r^2\alpha_n^2\right]^{1/2}
   \right|_{l=0}^{\infty}+O\left(\frac{f}{r\kappa n}\right).
\end{align}
Therefore, for $n\ge 1$,
\begin{align}
I_n&=\frac{1}{r\sqrt{f}}\left\{-\left[\frac{1}{4}+r^2
\left(m^2+\frac{r^2\kappa ^2n^2}{f^2}\right)\right]^{1/2}
+O\left(\frac{f}{r\kappa n}\right)\right\}\notag \\
  &=-\frac{\kappa n}{f}-\frac{m^2}{2\kappa n}+O\left(\frac{f}{r
n^3},\frac{\sqrt{f}}{r^2\kappa n}\right).
\end{align}
By taking $n$ sum, $O(f/(rn^3),\sqrt{f}/(r^2\kappa n))$ terms give finite
values which becomes zero on the horizon and we can omit these terms. Using
the formula
\begin{equation}
\begin{split}
 \kappa\sum_{n=1}^{\infty}\frac{\cos(n\kappa\ep)}{n\kappa}&=-\frac{1}{2}\ln(
\kappa^2\ep^2)+O(\ep^2),\\
 \kappa\sum_{n=1}^{\infty}n\kappa\cos(n\kappa\ep)&=-\frac{1}{\ep^2}
-\frac{\kappa ^2}{12}+O(\ep^2),
\end{split}
\end{equation}
we have
\begin{align}
  \langle\phi^2\rangle_{n\ge 1}&\approx -\frac{\kappa^2}{4\pi^2
f}\sum_{n=1}^{\infty}n\cos(n\kappa\ep)-\frac{m^2}{8\pi^2}\sum_{n=1}^{\infty}
\frac{\cos(n\kappa\ep)}{n}\notag \\
 &=\frac{1}{4\pi^2f\ep^2}+\frac{\kappa^2}{48\pi^2f}+\frac{m^2}{16\pi^2}\ln(\
kappa^2\ep^2).
\end{align}
Using the DeWitt-Schwinger counter term (\ref{eq:DS}), the regularized
expression of the expectation value becomes
\begin{equation}
 \langle\phi^2\rangle_{n\ge 1}-\langle\phi^2\rangle_{DS}=
  -
  \frac{m^2}{16\pi ^2}\ln \left(\frac{m^2 f}{16 \pi ^2 T^2} \right)
  +\frac{m^2}{16\pi ^2} +\frac{T^2}{12f} -
  \frac{{f'}^2}{192\pi^2 f}-\frac{f''}{96\pi ^2}-\frac{f'}{48\pi ^2r}.
  \end{equation}
This gives the same expression  of
Anderson {\it et al.}\cite{Anderson3}, who used the second order WKB
solution. Near the black hole horizon, it is sufficient to use the zeroth
order WKB solution to reproduce the result of the second order WKB
approximation. On the horizon $f=0$, this expression diverge and we must
examine the contribution from $n=0$ mode.

\subsection{$n=0$ contribution}

We must recall that the expression (\ref{eq:nge 1mode}) does not
include $n=0$ mode but there is no reason to exclude this contribution
in the case of the massive scalar field. So we must investigate the
contribution of $n=0$ mode. Since the WKB method breaks down for $n=0$
mode near the horizon, we cannot apply the approximation used by
Anderson \textit{et al.}\cite{Anderson3}. We solve $n=0$ mode function by
the following
method.

As we assume that the mass of the scalar field is large, a dimensionless
constant
\begin{equation}
  \ep\equiv\frac{r_+-r_-}{4 r_+[m^2r_+^2+l(l+1)]}
\end{equation}
becomes smaller than unity for all value of $l$:
\begin{equation}
  \ep\ll 1\qquad (\text{for all } l)
\end{equation}
We rescale the radial coordinate $r$ as follows:
\begin{equation}
  \label{eq:}
  x\equiv \sqrt{\frac{r-r_+}{\ep r_+}}.
\end{equation}
Assuming that $x$ is $O(1)$ quantity, this rescaling of the radial
coordinate means we are concentrate on the region $(r-r_{+})\sim
O(\ep)$, which is the region near the black hole horizon.  Using $x$
and $\ep$, the radial equation (\ref{eq:modeeq}) for $n=0$ mode
can be written as
\begin{eqnarray}
  \label{eq:modeeq2}
  \frac{d^2 \chi _{0l}}{dx^2} +\left[\frac{1}{x}+F(x,\ep)\right]
  \frac{d \chi _{0l}}{dx} -\left[1+ G(x,\ep)\right]\chi _{0l}=0,
\end{eqnarray}
where
\begin{align}
  F(x,\ep)&=\frac{2\ep x}{2\kappa r_{+}+\ep x^2},\\
  G(x,\ep)&=\frac{16\ep}{2\kappa r_{+}+\ep x^2}
  \left[\kappa r_{+}+2m^2r_{+}^2(2\ep^2 x^2+3\ep^3x^4)\right].
\end{align}
We can express the approximate solution of the mode equation
(\ref{eq:modeeq2}) by the power series expansion with respect to a
small parameter $\ep $:
\begin{equation}
  \label{eq:modesol}
  \chi_{0l}(x, \eta)=\chi_{0l}^{(0)}(x)+\ep\,\chi_{0l}^{(1)}(x)
  +\ep^2\,\chi_{0l}^{(2)}(x)+\cdots.
\end{equation}
$\chi_{0l}^{(n)}$ obeys
\begin{align}
  &\frac{d\chi_{0l}^{(0)}}{dx^2}+\frac{1}{x}\frac{d\chi_{0l}^{(0)}}{dx}
  - \chi_{0l}^{(0)}=0, \\
  &\frac{d\chi_{0}^{(n)}}{dx^2}+\frac{1}{x}\frac{d\chi_{0l}^{(n)}}{dx}
  - \chi_{0l}^{(n)}=  \notag \\
  & \qquad\qquad
\sum_{k=1}^n\frac{1}{k!}\left[-\left.\frac{\partial^kF}{\partial\ep^k}\right
|_{\ep=0}\frac{d\chi_{0l}^{(n-k)}}{dx}+\left.\frac{\partial^kG}{\partial\ep^
k}\right|_{\ep=0}\chi_{0l}^{(n-k)}\right]
  \qquad (\text{for }n\ge1)
\end{align}
We can obtain solutions of these equations by using modified Bessel
functions. For example, the lowest order solution is given by
\begin{equation}
 p_{l0}^{(0)}=I_0(x),\qquad q_{l0}^{(0)}=\frac{2}{r_{+}-r_{-}}K_0(x).
\end{equation}
 Substituting the expansion (\ref{eq:modesol}) to
\begin{equation}
 \langle\phi^2\rangle_{n=0}=\frac{T}{4\pi}\sum_{l=0}^{\infty}\left[(2l+1)p_{
l0}q_{l0}-\frac{1}{r\sqrt{f}}\right]
\end{equation}
and taking summation about $l$ using the Plana sum
formula, $\langle\phi^2\rangle_{n=0}$ is written as the
power series expansion with respect to $(m r_{+})^{-1}$. We calculated
$\langle\phi^2\rangle_{n=0}$ up to $O\left((mr_{+})^{-4}\right)$, which is
accomplished by the $\ep$ expansion of the mode function up to
$O(\ep^3)$. By taking into the account of the contribution of $n=0$
mode, we finally get the form of the vacuum polarization near the
horizon as follows:
\begin{align}
  \label{eq:results}
  \langle \phi ^2 \rangle _{T} & =
  \left(T-\frac{\kappa}{2\pi}\right)
  \left\{-\frac{m^2}{16\pi \kappa} +\frac{m^2}{8\pi \kappa}
    \left[2\gamma +\ln\frac{m^2(r-r_+)}{2\pi T}\right] -\frac{1}{12\pi
      r_+}\right\} \notag \\
  &+\left\{T^2-\left(\frac{\kappa}{2\pi}\right)^2\right\}
  \left[\frac{1}{24\kappa(r-r_+)}-\frac{r_-}{48r_+^3\kappa ^2}\right]
  \nonumber \\
  &  +\frac{m^2}{16\pi ^2 \kappa}(\kappa+2\pi T) \ln
  \left(\frac{2\pi T}{\kappa}\right)\notag \\
  & +\frac{T}{4\pi }
  \frac{1}{45 m^2}\frac{1}{(r_+-r_-)r_{+}^2}
  \left[3-\frac{6r_-}{r_+}+\frac{4r_-^2}{r_+^2} \right]\notag \\
  &+\frac{T}{4\pi }\frac{2}{315 m^4}\frac{1}{(r_+-r_-)r_{+}^4}
  \left[-10+39\frac{r_-}{r_+}-52\frac{r_-^2}{r_+^2}
    +23\frac{r_-^3}{r_+^3}\right]+O\left((mr_{+})^{-6}\right) .
\end{align}
The divergences in $\langle \phi ^2 \rangle_T$ appears only in the
terms of $O(m^{2})$ and $O(m^{0})$.  Taking $f\rightarrow 0$ limit,
$O(m^2)$ term diverge as $\ln(r-r_{+})$ and $O(m^0)$ as
$(r-r_{+})^{-1}$. These divergence disappear completely if the
temperature of the scalar field equals $T= \kappa /2\pi \equiv T_H$,
where $T_H$ is the Hawking temperature.  Specifying $T=T_H $ is
equivalent to require the regurality of $\langle \phi ^2 \rangle$ on
the horizon. In this case, the black hole is equilibrium with the
thermal scalar field with temperature $T=T_H$. This is nothing but a
Hartle-Hawking state.

We can evaluate a finite part of $\langle \phi ^2 \rangle_T $ in the
 Hartle-Hawking state on the horizon.  The leading order of $\langle \phi ^2
 \rangle_T $ at $T=T_H$ is $O(m^{-2})$ because  not only divergent but
 also finite terms contained in $O(m^2)$
and $O(m^0)$ terms vanish. The result is given by
\begin{equation}
\begin{split}
  \label{eq:massive}
  \langle \phi ^2 \rangle _{T= T_H}&=\frac{1}{720 \pi ^2
    m^2}\frac{1}{r_{+}^4}
  \left[3-\frac{6r_-}{r_+}+\frac{4r_-^2}{r_+^2} \right] \\
 &
  +\frac{1}{2520 \pi ^2 m^4}\frac{1}{r_{+}^6}
  \left[-10+39\frac{r_-}{r_+}-52\frac{r_-^2}{r_+^2}
    +23\frac{r_-^3}{r_+^3}\right]+O\left(m^{-6}\right).
\end{split}
\end{equation}
This expression agrees with the result of the DeWitt-Schwinger expansion
\cite{Frolov,Parker,Anderson1}.

\section{Conclusion and discussion }
We calculated $\langle \phi ^2 \rangle_T$ of a massive scalar field
near the horizon of a Reissner-Nordstr\"{o}m black hole using point
splitting method.  Our results are as follows: (i)all divergent and
finite terms of $O(m^{2})$ and $O(m^{0})$ in $\langle \phi ^2
\rangle_T$ are canceled to be zero if the temperature of the scalar
field equals to the Hawking temperature $T_H$ of a black hole.
(ii)at this temperature, the leading order of the regularized $\langle
\phi ^2 \rangle_T$ becomes $O(m^{-2})$ and its value on the horizon
agrees with that of the DeWitt-Schwinger expansion up to $O(m^{-4})$.
(i) and (ii) indicate that regularized $\langle \phi ^2 \rangle_T$ in
the Hartle-Hawking state near the black hole horizon is well
approximated by the DeWitt-Schwinger expansion. This confirms the
results of \cite{Frolov,Frolov2}.

It is not trivial that the finite part of $\langle \phi ^2 \rangle_T$
in the thermal state at $T=T_H$ is reproduced by the DeWitt-Schwinger
expansion which does not specify any thermal state of the scalar field. It
was
shown that the leading order of the DeWitt-Schwinger expansion
reproduces the value $\langle\phi^2\rangle_{T=0}$ of scalar field with
large mass in general static spherical symmetric spacetime
\cite{Herman}.  This result is natural because the scalar field with $T=0$
is in a vacuum state and  the
DeWitt-Schwinger expansion does not incorporate
information on a thermal state.

Discrepancy between $\langle \phi ^2 \rangle_T$
in the Hartle-Hawking state and that in he DeWitt-Schwinger expansion will
appear far
from the horizon, because the thermal effect dominates the
gravitational effect in the asymptotic flat region. If the thermal
effect is dominant or comparable to the contribution from curvature
effect near the horizon, there would also be discrepancy between $\langle
\phi ^2 \rangle_T$ in the Hartle-Hawking state and that of the
DeWitt-Schwinger expansion.  However this is not seen from
our results. This implies that the finite part of $\langle \phi^2
\rangle_T$ in the Hartle-Hawking state near the black hole horizon is
dominated by the contribution from curvature effect.  Why the
contribution from thermal effects to finite parts can be negligible?
  The characteristic scale for
thermal state is the inverse of the temperature $1/T_H \sim M$ which
is a scale length of thermal fluctuation.  On the other hand, the
Compton length of the scalar field is $1/m$. The contribution of the
Hawking radiation to $\langle\phi^2\rangle$ is $\propto T_H^4/m^2$ and
that of thermal excitation is $\propto\exp(-m/T_H)$. For $m\gg 1/M$,
the effect of thermal excitation is exponentially small. Furthermore,
near the black hole horizon, almost all part of thermally excited
particle is absorbed by the black hole. This is the reason of the
DeWitt-Schwinger approximation gives the same result of Hartle-Hawking
state on the horizon.




\end{document}